\documentclass[runningheads]{llncs}

\usepackage{graphicx}
\usepackage{epstopdf}
\usepackage{algorithm}
\usepackage{algorithmic}
\usepackage{mathtools}
\usepackage{amsmath,amssymb}
\usepackage{setspace}
\usepackage{enumerate}
\usepackage{multirow}
\usepackage{float}
\usepackage{array}
\usepackage{booktabs}
\usepackage{threeparttable}
\usepackage{cite}
\usepackage{color,array}

\usepackage{makecell}

\begin{document}

\title{Protect the Intellectual Property of Dataset against Unauthorized Use}

\author{Mingfu Xue\inst{1} \and
Yinghao Wu\inst{1} \and
Yushu Zhang\inst{1} \and
Jian Wang\inst{1} \and
Weiqiang Liu\inst{2}}

\authorrunning{M. Xue et al.}

\institute{College of Computer Science and Technology, Nanjing University of Aeronautics and Astronautics, Nanjing, China \\
\and
College of Electronic and Information Engineering, Nanjing University of Aeronautics and Astronautics, Nanjing, China}

\maketitle
\begin{abstract}
Training high performance Deep Neural Networks (DNNs) models require large-scale and high-quality datasets.
The expensive cost of collecting and annotating large-scale datasets make the valuable datasets can be considered as the Intellectual Property (IP) of the dataset owner.
To date, almost all the copyright protection schemes for deep learning focus on the copyright protection of models, while the copyright protection of the dataset is rarely studied.
In this paper, we propose a novel method to actively protect the dataset from being used to train DNN models without authorization.
Experimental results on on CIFAR-10 and TinyImageNet datasets demonstrate the effectiveness of the proposed method.
Compared with the model trained on clean dataset, the proposed method can effectively make the test accuracy of the unauthorized model trained on protected dataset drop from 86.21\% to 38.23\% and from 74.00\% to 16.20\% on CIFAR-10 and TinyImageNet datasets, respectively.

\end{abstract}

\section{Introduction}
The performance of the deep neural networks greatly depends on the quantity and quality of the training data.
Therefore, datasets, especially large-scale and high-quality datasets, have tremendous impact on training a high performance Deep Neural Networks (DNN).
However, the process of collection, verification, and annotation of large-scale dataset is time-consuming and expensive, where the images need to be manually labelled by experienced experts.
As a consequence, high-quality and large-scale datasets can be considered as the Intellectual Property (IP) of the dataset owner.
To date, many researches have been proposed to protect the Intellectual Property (IP) of the Deep Neural Networks \cite{UchidaNSS17, NagaiUSS18, AdiBCPK18, ZhangGJWSHM18} models.
However, the copyright protection of the dataset is rarely studied.
In this paper, we propose a novel method to actively protect the dataset from unauthorized use.
The clean dataset is transformed to the protected dataset.
For unauthorized users, the model trained on the protected dataset will show a very poor performance on test data.
In contrast, the authorized users can provide the protected dataset to the Dataset Management Cloud Platform (DMCP) to restore the ``clean'' dataset and train their model on the recovered dataset. The model trained on the recovered dataset will have a normal performance on test data.

This paper is organized as follows.
The proposed protection method is elaborated in Section \ref{sec:proposed_method}.
Experimental results are presented and discussed in Section \ref{sec:exp_results}.
This paper is concluded in Section \ref{sec:con}.

\section{The Proposed Method}
\label{sec:proposed_method}
\subsection{Overall Flow}
\begin{figure}[htbp]
\centering
\includegraphics[width=\textwidth]{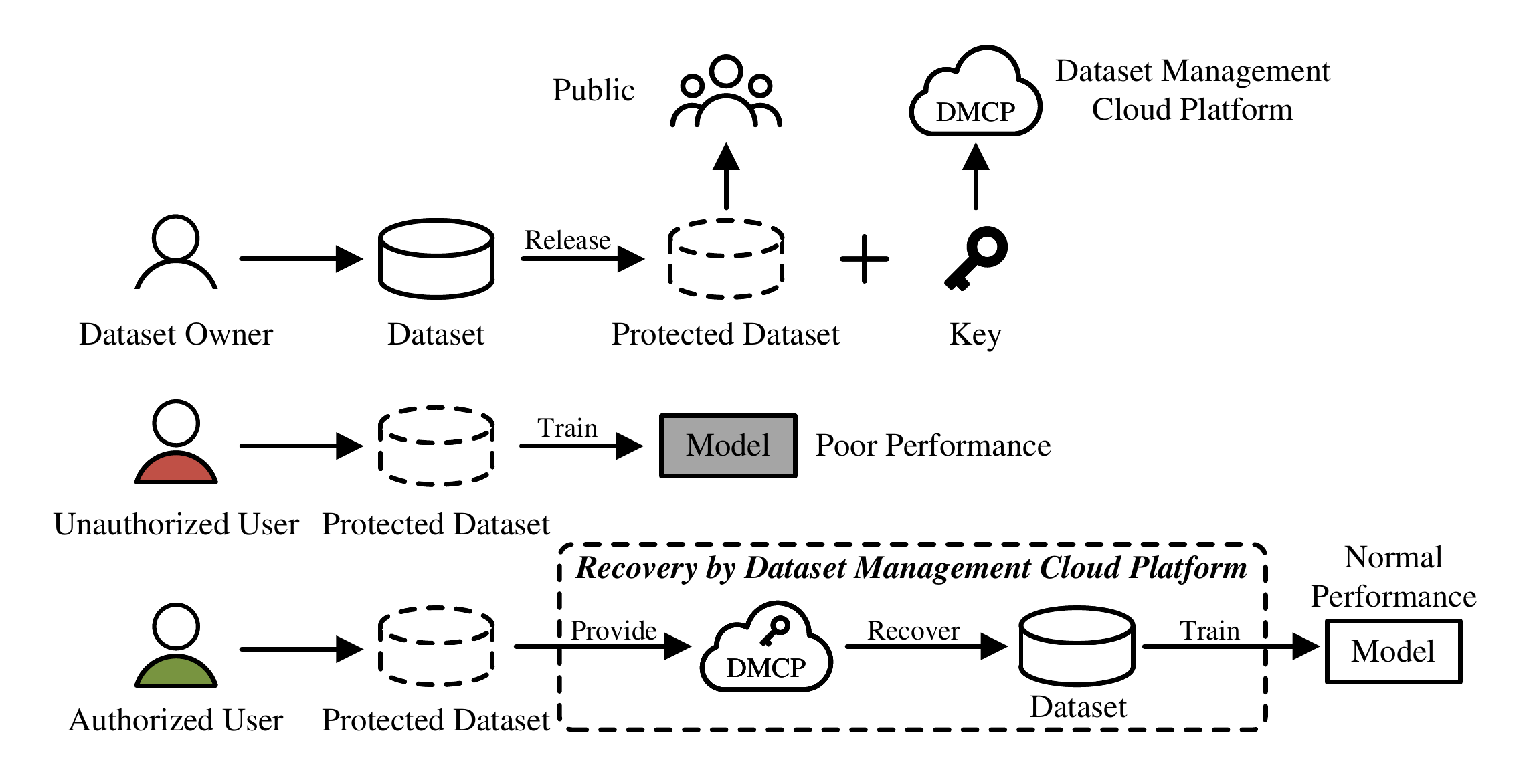}
\caption{The overall flow of the proposed method.}
\label{fig:overall}
\end{figure}

The overall flow of the proposed method is illustrated in Fig. \ref{fig:overall}.
The proposed method can be divided into three parts:
\begin{itemize}
\item The owner of the valuable dataset transform the clean dataset to the protected dataset through the proposed method.
Then, the protected dataset is released to the public and the secret key is safely kept in the Dataset Management Cloud Platform.
\item For an unauthorized user, the protected dataset is directly used to train the model.
The inference performance of the trained model will be significantly deteriorated due to the protection of the proposed method.
\item For an authorized user, he can provide the protected dataset (A subset of the dataset or the entire dataset) to the DMCP. The DMCP will restore the dataset with the secret key.
The inference performance of the model trained on the recovered dataset will be consistent to that of the model trained on clean dataset.
\end{itemize}

\begin{figure}[htbp]
\centering
\includegraphics[width=\textwidth]{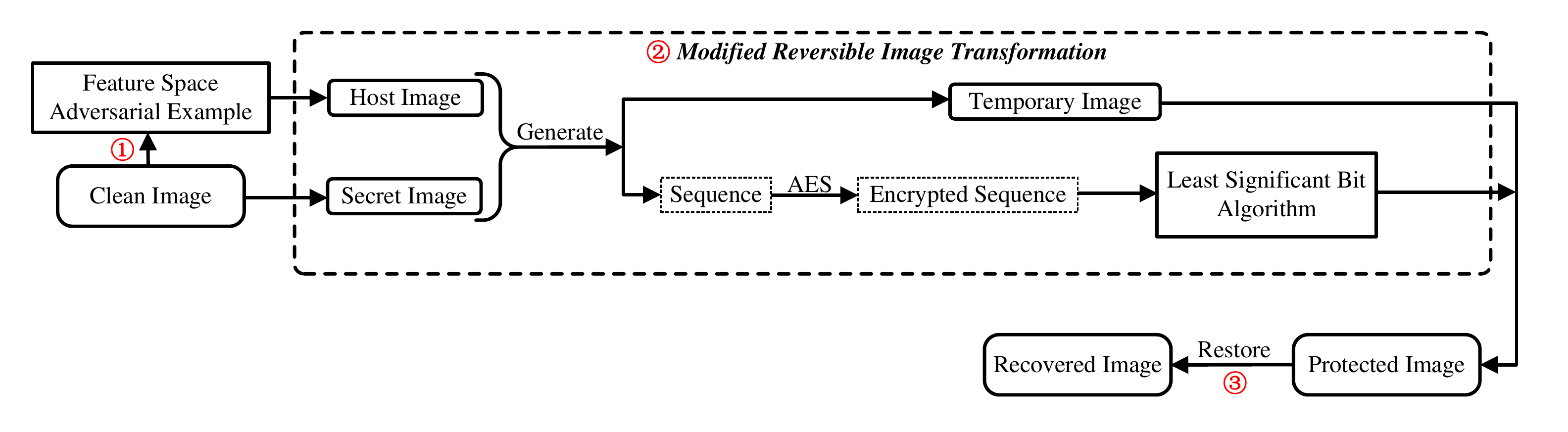}
\caption{The framework of the proposed method.}
\label{fig:method}
\end{figure}

\subsection{Proposed Method}
In this section, we elaborate the proposed method, which includes protected image generation and clean image restoration.
As shown in Fig. \ref{fig:method}, the framework includes three steps:
1) First, the clean image is perturbed through the feature space adversarial example generation method, which is proposed in work \cite{InkawhichWLC19}.
2) Second, we exploit a modified Reversible Image Transformation (mRIT) method to hide the clean image into the perturbed image to generate the protected image.
The mRIT method is modified from the Reversible Image Transformation (RIT) method \cite{ZhangWHY16}.
3) Lastly, the ``clean'' image can be recovered from the protected image with a secret key on the Dataset Management Cloud Platform.
This recovery process is inverse to the process of mRIT.
The above three steps are described as follows:

\begin{itemize}
\item \textbf{Feature Space Adversarial Example Generation.}
The perturbed image is generated through perturbing the clean image using the Feature Space Perturbation (FSP) \cite{InkawhichWLC19}.
Adversarial examples generated by adding feature space perturbations are more transferable than that generated through gradient-based or optimization-based method \cite{InkawhichWLC19}.
\item \textbf{Modified Reversible Image Transformation (mRIT).}
First, the clean image is embedded into the perturbed image through block pairing and block transformation to generate the temporary image.
The method of block pairing and block transformation is proposed in work \cite{ZhangWHY16}.
Second, in order to perform image restoration, the \textit{Sequence} that contains the information of block pairing and transformation needs to be be encrypted and embedded into the temporary image.
Specifically, we encrypt the \textit{Sequence} by a standard encryption scheme (AES) \cite{nechvatal2001report} with a secret key.
Then, we leverage Least Significant Bit (LSB) algorithm \cite{lsbsteganography2020} to embed the encrypted \textit{Sequence} into the temporary image to obtain the final protected image.
\item \textbf{Image Restoration.}
The image restoration process is the reverse process of the mRIT.
First, the encrypted \textit{Sequence} is extracted from the protected image.
Second, the encrypted \textit{Sequence} is decrypted with the secret key.
Lastly, the ``clean'' image is restored from the protected image with the \textit{Sequence}.
\end{itemize}

\section{Experimental Results}
\label{sec:exp_results}
\begin{figure}[htbp]
\centering
\includegraphics[scale=0.9]{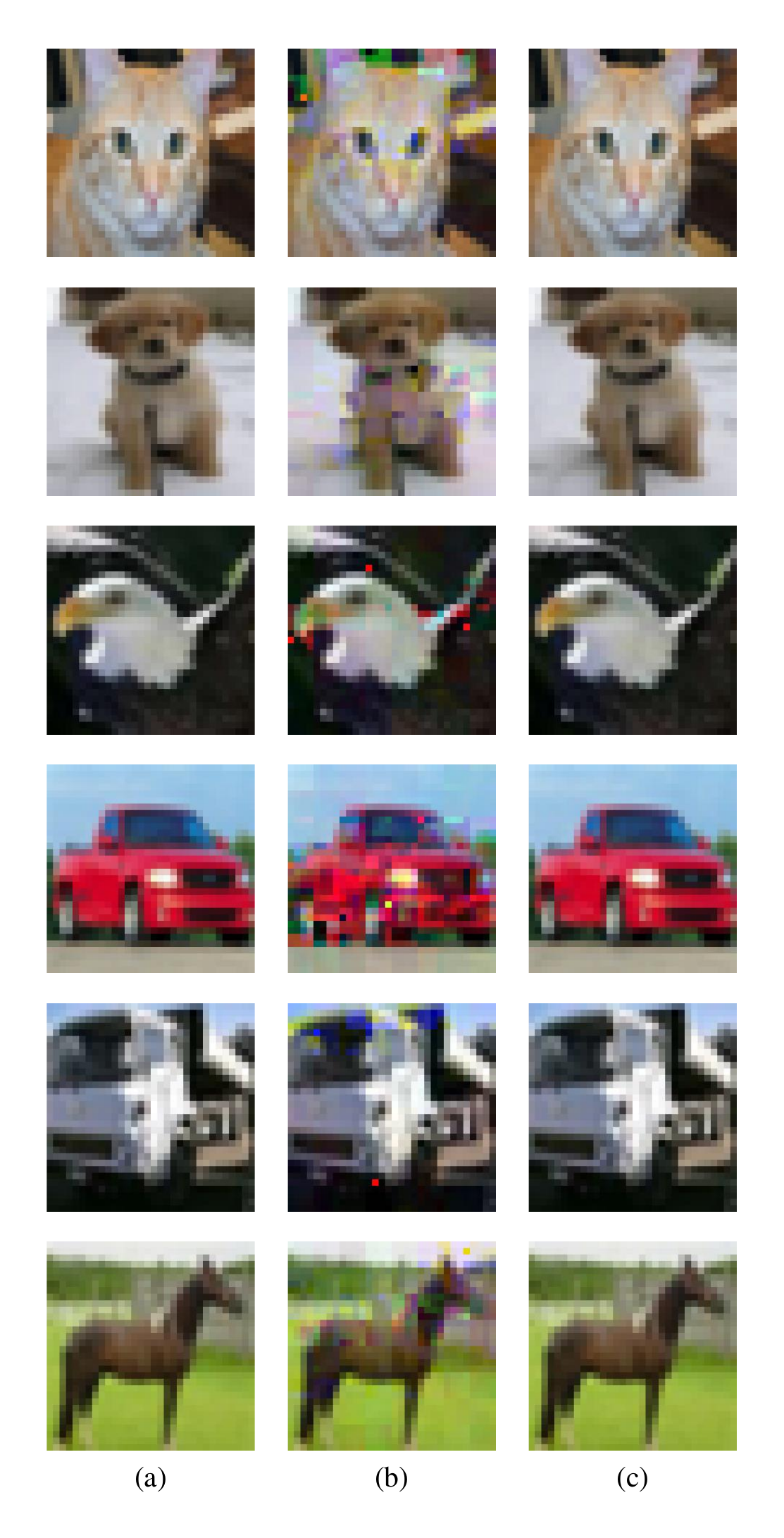}
\caption{Examples of clean images, protected images and recovered images of CIFAR-10 dataset: (a) Clean Images, (b) Protected Images, (c) Recovered Images}
\label{fig:trans_cifar10}
\end{figure}

\begin{figure}[htbp]
\centering
\includegraphics[scale=0.9]{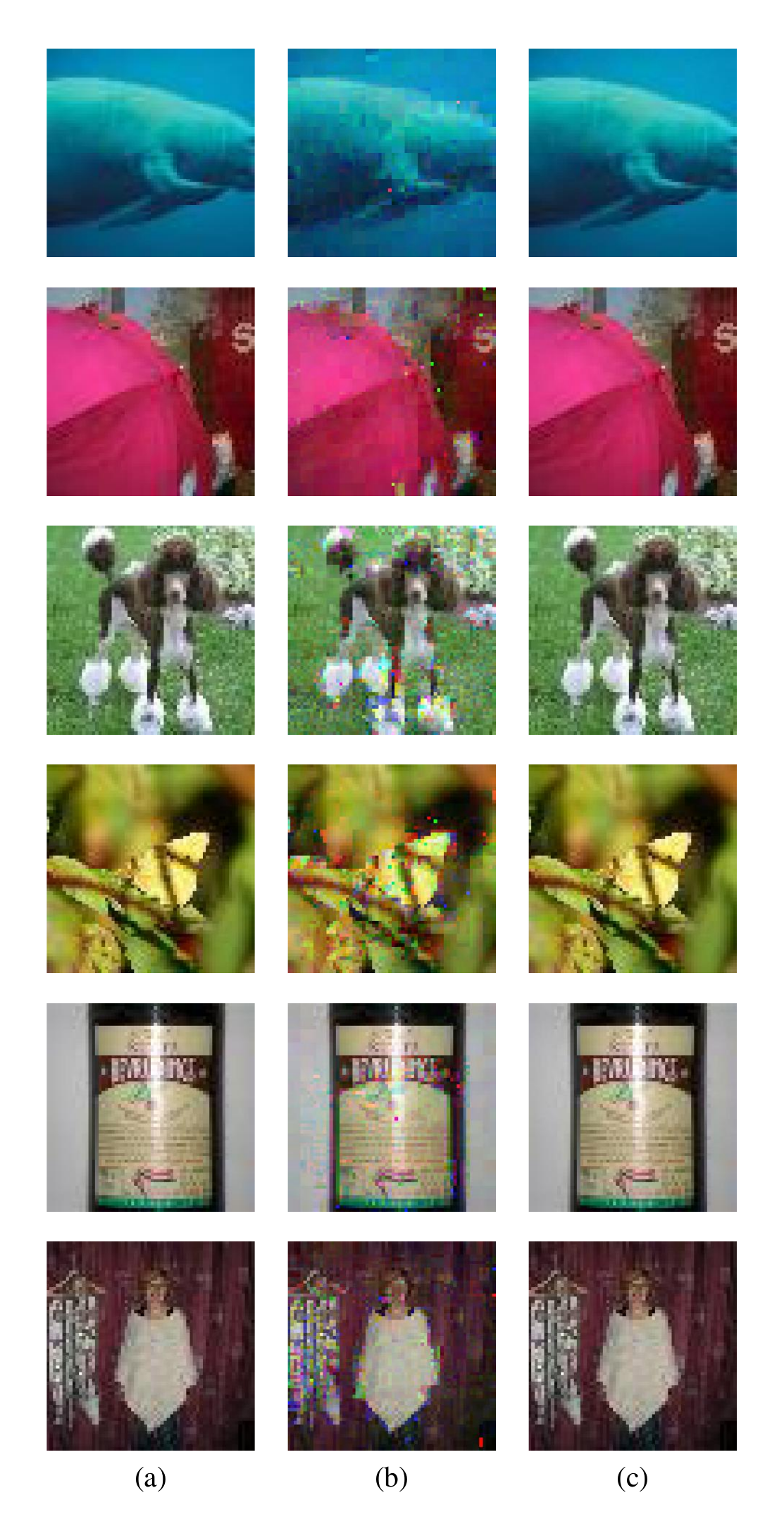}
\caption{Examples of clean images, protected images and recovered images of TinyImageNet dataset: (a) Clean Images, (b) Protected Images, (c) Recovered Images}
\label{fig:trans_tinyimagenet}
\end{figure}

Experiments are conducted on the CIFAR-10 \cite{Krizhevsky_2009_17719} and TinyImageNet \cite{tinyimagenet200} datasets.
The model architecture is ResNet-18 \cite{HeZRS16}.
For CIFAR-10 dataset \cite{Krizhevsky_2009_17719}, there are 10 classes.
Each class has 5,000 training images and 1,000 test images.
In TinyImageNet \cite{tinyimagenet200} dataset, there are 500 training images and 50 test images for each class, and we randomly sample 10 classes of images from TinyImageNet.
All the training images of CIFAR-10 and the above subset of TinyImageNet are used as the clean training images.
The corresponding protected images are generated through the proposed method.
The ``clean'' training images are recovered from the protected images correspondingly.
As shown in Fig. \ref{fig:trans_cifar10} and Fig. \ref{fig:trans_tinyimagenet}, the images at left, middle, right column are the clean images, protected images, and recovered images respectively.
The ResNet-18 \cite{HeZRS16} model is trained from scratch on clean, protected and reversed dataset, respectively.
For the training process of ResNet-18 on CIFAR-10, we utilize Adam optimizer \cite{KingmaB14} and set the initial learning rate as 0.01.
The batch size and training epoch are set to be 128 and 80, respectively.
The experimental settings on TinyImageNet are the same as the settings on CIFAR-10 dataset. All the test images of the CIFAR-10 and the above subset of TinyImageNet dataset are used to evaluate the test accuracy of the trained models.

The experimental results are shown in Table. \ref{tab:exp_rst}.
As shown in Table. \ref{tab:exp_rst}, the inference performance of the model trained on clean dataset is normal.
Specifically, the test accuracy of the model on clean CIFAR-10 \cite{Krizhevsky_2009_17719} and TinyImageNet \cite{tinyimagenet200} is 86.21\% and 74.00\%, respectively.
However, the inference performance of the model trained on the dataset under the protection of the proposed method significantly degrades.
Specifically, the test accuracy of the model trained on protected CIFAR-10 and TinyImageNet has drastically reduced to 38.23\% and 16.20\% respectively, which demonstrates that the illegitimate usage of the dataset to train DNN model will cause the trained model to have a very poor performance on test data.
Lastly, the inference performance of the model trained on recovered dataset is consistent with that of the model trained on clean dataset.
Specifically, the test accuracy of the model on recovered CIFAR-10 \cite{Krizhevsky_2009_17719} and TinyImageNet \cite{tinyimagenet200} is 85.86\% and 73.20\% respectively, which demonstrates that the performance of the model on recovered dataset is not affected.

\begin{table}[htbp]
  \centering
  \caption{The training accuracy and test accuracy of the ResNet-18 model on clean, protected and recovered dataset of CIFAR-10 and TinyImageNet.}
    \begin{tabular}{|c|c|c|c|c|}
    \hline
    Model & \multicolumn{2}{c|}{Dataset} & Training Accuracy & Test Accuracy \\
    \hline
    \multirow{6}*{ResNet-18} & \multirow{3}*{CIFAR-10} & Clean & 96.44\% & 86.21\% \\
\cline{3-5}          &       & Protected & 100.00\% & 38.23\% \\
\cline{3-5}          &       & Reversed & 95.21\% & 85.86\% \\
\cline{2-5}          & \multirow{3}*{TinyImageNet} & Clean & 99.98\% & 74.00\% \\
\cline{3-5}          &       & Protected & 100.00\% & 16.20\% \\
\cline{3-5}          &       & Reversed & 99.96\% & 73.20\% \\
    \hline
    \end{tabular}%
  \label{tab:exp_rst}%
\end{table}%

\section{Conclusion}
\label{sec:con}
In this paper, we propose a novel method to protect the Intellectual Property of valuable datasets from being used without authorization.
With the proposed protection method, the model trained on the protected dataset will have a very poor performance on test data.
Experimental results on CIFAR-10 \cite{Krizhevsky_2009_17719} and TinyImageNet \cite{tinyimagenet200} datasets reveal that the proposed dataset protection method can effectively make the model trained on the protected dataset without authorization have a poor performance while the authorized users can obtain high performance with recovered clean images.

\bibliographystyle{splncs04}
\bibliography{ref}

\begin{thebibliography}{10}
\providecommand{\url}[1]{\texttt{#1}}
\providecommand{\urlprefix}{URL }
\providecommand{\doi}[1]{https://doi.org/#1}

\bibitem{AdiBCPK18}
Adi, Y., Baum, C., Ciss{\'{e}}, M., Pinkas, B., Keshet, J.: Turning your
  weakness into a strength: Watermarking deep neural networks by backdooring.
  In: 27th {USENIX} Security Symposium, {USENIX}. pp. 1615--1631 (2018)

\bibitem{lsbsteganography2020}
David, R.: {LSB-S}teganography.
  \url{https://github.com/RobinDavid/LSB-Steganography} (2020)

\bibitem{HeZRS16}
He, K., Zhang, X., Ren, S., Sun, J.: Deep residual learning for image
  recognition. In: {IEEE} Conference on Computer Vision and Pattern
  Recognition, {CVPR}. pp. 770--778 (2016)

\bibitem{InkawhichWLC19}
Inkawhich, N., Wen, W., Li, H.H., Chen, Y.: Feature space perturbations yield
  more transferable adversarial examples. In: {IEEE} Conference on Computer
  Vision and Pattern Recognition, {CVPR}. pp. 7066--7074 (2019)

\bibitem{tinyimagenet200}
Karpathy, A.: {Cs231n}: {Convolutional} neural networks for visual recognition.
  Neural Networks  \textbf{1}(1), ~1--1 (2016)

\bibitem{KingmaB14}
Kingma, D.P., Ba, J.: Adam: {A} method for stochastic optimization. In: 3rd
  International Conference on Learning Representations, {ICLR}. pp. 1--15
  (2015)

\bibitem{Krizhevsky_2009_17719}
Krizhevsky, A., Hinton, G.: Learning multiple layers of features from tiny
  images. Tech. rep., University of Toronto (2009)

\bibitem{NagaiUSS18}
Nagai, Y., Uchida, Y., Sakazawa, S., Satoh, S.: Digital watermarking for deep
  neural networks. Int. J. Multim. Inf. Retr.  \textbf{7}(1),  3--16 (2018)

\bibitem{nechvatal2001report}
Nechvatal, J., Barker, E., Bassham, L., Burr, W., Dworkin, M., Foti, J.,
  Roback, E.: Report on the development of the advanced encryption standard
  (aes). Journal of Research of the National Institute of Standards and
  Technology  \textbf{106}(3), ~511 (2001)

\bibitem{UchidaNSS17}
Uchida, Y., Nagai, Y., Sakazawa, S., Satoh, S.: Embedding watermarks into deep
  neural networks. In: Proceedings of the 2017 {ACM} on International
  Conference on Multimedia Retrieval, {ICMR}. pp. 269--277 (2017)

\bibitem{ZhangGJWSHM18}
Zhang, J., Gu, Z., Jang, J., Wu, H., Stoecklin, M.P., Huang, H., Molloy, I.M.:
  Protecting intellectual property of deep neural networks with watermarking.
  In: Proceedings of the 2018 on Asia Conference on Computer and Communications
  Security. pp. 159--172 (2018)

\bibitem{ZhangWHY16}
Zhang, W., Wang, H., Hou, D., Yu, N.: Reversible data hiding in encrypted
  images by reversible image transformation. {IEEE} Trans. Multim.
  \textbf{18}(8),  1469--1479 (2016)

\end{thebibliography}

\end{document}